\documentclass{webofc}
\usepackage[varg]{txfonts}   
\usepackage[font=small,labelfont=bf,justification=justified,format=plain ]{caption}
\usepackage{color}
\definecolor{purple}{rgb}{0.5,0,0.5}
\definecolor{blue}{rgb}{0.0,0,0.9}
\usepackage[colorlinks=true, pdfstartview=FitV, citecolor= purple, linkcolor = blue,urlcolor=blue]{hyperref}

\begin{document}

\vspace*{-2cm}

\title{The Charm and Beauty of Strong Interactions}

\author{\firstname{Bruno} \lastname{El-Bennich}\inst{1} \fnsep\thanks{\email{bruno.bennich@cruzeirodosul.edu.br}} }

\institute{Laborat\'orio de F\'isica Te\'orica e Computacional, Universidade Cruzeiro do Sul, Rua Galv\~ao Bueno 868, 01506-000 S\~ao Paulo, SP, Brazil }

\abstract{We briefly review common features and overlapping issues in hadron and flavor physics focussing on continuum QCD approaches to heavy
bound states, their mass spectrum and weak decay constants in different strong interaction models.}
\maketitle

\section{ Hadron and Flavor Physics: a twofold motivation, a unified approach? }

Hadron and flavor physics are often thought of as different research fields though they have in common the underlying theory which describes strong interactions in flavored 
and non-flavored hadrons, namely Quantum Chromodynamics (QCD). In particular {\em confinement\/}, one of QCD's paradigmatic hallmarks, may not be appreciated enough 
in flavor physics. Undoubtedly, in nonrelativistic QCD and effective field theory approaches to quarkonia and the fauna of new $XYZ$ states the main focus is on spectroscopy 
and production mechanisms, where most efforts concentrate on nonrelativistic perturbative aspects, effective theories or potentials models~\cite{Bodwin:1994jh,Brambilla:2010cs,
Brambilla:2015rqa,Brambilla:2017uyf}.

While certain quarkonia provide a relatively clean environment to probe strong dynamics, and the mass spectrum of their confined quarks is reasonably well described in nonrelativistic
perturbative QCD (pQCD), many of the heavier $XYZ$ states close to threshold may not be pure $\bar qq$ states but admixtures of meson molecules and a core of quarks. Moreover, it is 
desirable  to unify the meson {\em and\/} baryon spectrum with a unique confining framework that correctly describes the dynamical chiral symmetry breaking (DCSB) pattern which occurs to 
varying degrees in hadrons. In particular, DCSB may still play an important role in charmonia where nonperturbative $\mathcal{O}(\Lambda_\mathrm{QCD}/m_c)$  contributions can be 
non negligible.  On the other hand, the study of weak $B$-meson decays, once driven by factorization theorems~\cite{Beneke:1999br} and soft collinear effective field theory~\cite{Bauer:2000yr}, 
appears to be nowadays dominated by the quest for new physics largely motivated by recurrent announcements of the LHC$b$ collaboration~\cite{Aaij:2015oid,Abdesselam:2016llu,Aaij:2017vbb}. 
Still, it is exactly in the case of heavy-light mesons where our understanding of their QCD dynamics and emerging properties are the least well understood within continuum QCD
approaches~\cite{ElBennich:2006yi,Leitner:2010fq,ElBennich:2010ha,ElBennich:2011py,El-Bennich:2013yna,Rojas:2014aka,El-Bennich:2016bno,Mojica:2017tvh,Gomez-Rocha:2015qga,
Hilger:2017jti}. The reason for this is obviously the important difference in quark  masses and the energetic light mesons  produced in their decays, both of which are at the origin of a wide 
array of energy scales~\cite{ElBennich:2009vx,ElBennich:2012tp} which effective theories can merely separate.

More generally, the main object of study in flavor physics is the origin of $CP$-violating phases in weak decays of flavored quarks {\em confined\/} in hadrons, where  the 
Cabibbo-Kobayashi-Maskawa (CKM) mechanism has been established as the dominant Standard Model source of $CP$ violation in heavy-meson decays.  On the other hand, the 
overwhelming bulk of visible matter is made of baryons containing light quarks. DCSB due to gluon self-interactions has by now been established to be the most important 
mass generating mechanism for visible matter in the Universe, which implies that for most observed matter the Higgs mechanism is almost irrelevant. Thus, both field's major 
preoccupation is the origin and generation of matter and pose the fundamental questions: {\em What is the origin of matter and its preponderance over antimatter? What is the 
origin of the mass of baryonic matter?\/} 

While the first question may not be completely answered within the Standard Model, as the CKM mechanism in the quark sector is not sufficient to explain our matter-dominated 
Universe~\cite{Quinn:2002yc}, the origin of mass due to DCSB has by now been firmly established~\cite{Cloet:2013jya,Eichmann:2016yit,Bali:2016lvx}. Moreover, there are good reasons 
to believe DCSB  and confinement are intimately related~\cite{Bashir:2012fs,OMalley:2011aa,Brambilla:2014jmp}. Experimental data ought to teach us something about the  interaction 
that holds quark-antiquark bound states together in different constellations, namely in ground states and in radially excited or larger angular-momentum states.  Quarkonia and heavy-flavored 
mesons consist of strongly interacting $\bar qq$ pairs that offer different perspectives on confinement: in the former case one tests QCD's running coupling at much shorter distances than 
for lighter mesons whose mass is chiefly due to DCSB, whereas in the latter case the same interaction is scrutinized in an asymmetric system giving rise to disparate energy scales 
where truncation effects are more evident~\cite{El-Bennich:2016qmb}.

At any rate, in calculations of hadronic observables a reliable description of DCSB and confinement cannot be realized with perturbative methods. Focusing on 
continuum QCD, so far the most successful approaches are based on QCD's equations of motion in Euclidean space, given by the infinite tower of coupled Dyson-Schwinger 
integral equations~\cite{Bashir:2012fs,Krein:1990sf,Fischer:2003rp}, and on the functional renormalization group~\cite{Pawlowski:2005xe}. We here review recent results on the
quarkonia and flavored meson mass spectrum and corresponding weak decay constants obtained in the combined framework of the Dyson-Schwinger equation (DSE) and Bethe-Salpeter 
equation (BSE) and stress the overlap of contemporary hadron physics studies, such as the experimental program of the 12~GeV upgrade at Jefferson Lab, with that of flavor 
physics facilities.

\section{Spectroscopy from continuum QCD}

Charmonia and Bottomonia have first been explored within a rainbow-ladder truncation of the gap and Bethe-Salpeter equations in the work by Jain and Munczek~\cite{Jain:1993qh}.
Driven by the ample growth of experimental data on quarkonia and in particular the discovery of many $XYZ$ states by $B$-factories, CLEO-$c$, BES II and BES III, more sophisticated 
BSE computations of the quarkonia spectra followed the initial steps and at least the masses of the lower $J^{PC}$ states compare very well with experimental data~\cite{Rojas:2014aka,
Mojica:2017tvh,Hilger:2017jti,Blank:2011ha,Fischer:2014cfa,Ding:2015rkn,Hilger:2014nma,Hilger:2015hka}.

Recently, the gap equation and BSE for heavier quarks have been revisited in the context of a given vector $\times$ vector contact interaction (CI), which introduces a mass scale 
and has zero range~\cite{Bedolla:2015mpa,Bedolla:2016yxq,Raya:2017ggu,Serna:2017nlr}. As for related Nambu-Jona-Lasinio models~\cite{Ninomiya:2017tke}, these CI models are not confining yet 
the implementation of a proper-time regularization scheme~\cite{Ebert:1996vx} prevents the quarks from going on their mass shell and hence emulates confinement. The spectrum produced 
with this interaction model compares quantitatively well with that obtained using more sophisticated finite-range QCD interaction models mentioned above. 

In a functional approach to QCD, the quark propagator for a given flavor $f$ is computed from the DSE,
\begin{equation}
S^{-1}_f(p)  =   Z_2^f \, (i\gamma\cdot p + m_f^\mathrm{bm} ) + Z_1^f\, g^2\!\!  \int^\Lambda_k \!\!  D^{\mu\nu} (k-p) \frac{\lambda^a}{2} \gamma_\mu\, S(k)\, \Gamma^f_\nu (k,p) \frac{\lambda^a}{2},
\label{DSE}
\end{equation}
where $\int_k^\Lambda$ represents a Poincar\'e invariant regularization of the integral and  $Z_{1,2}^f (\mu^2,\Lambda^2)$ are, respectively, the vertex and quark wave-function renormalization 
constants at the renormalization scale  $\mu$. Note that the regularization mass scale is much larger than the  renormalization scale: $\Lambda\gg \mu$. The current-quark mass, 
$Z_2^f (\mu,\Lambda ) m^f_\mathrm{bm} (\Lambda )$, receives self-energy corrections where the integral in the second term of Eq.~\eqref{DSE}  is over the dressed gluon propagator, 
$D_{\mu\nu}(k)$, the quark-gluon vertex, $\Gamma^f_\nu (k,p)$, and $\lambda^a$ are  the SU$(3)$ color matrices for quarks in the fundamental representation. The solution to the quarks's 
DSE~\eqref{DSE} contains a scalar and vector term,
\begin{equation}
  S(p)  =  -i \gamma\cdot p \ \sigma_V (p^2) + \sigma_S(p^2)  \equiv \left [ i \gamma\cdot p \  A(p^2) + B(p^2) \right ]^{-1}  .
  \label{sigmaSV}
\end{equation}
The full nonperturbative quark-gluon vertex, $\Gamma^a_\nu (k,p)$, plays a crucial role for DCSB~\cite{Aguilar:2016lbe,Aguilar:2010cn,Rojas:2013tza,Rojas:2014tya,Alkofer:2008tt,
Bashir:2011dp,Bermudez:2017bpx,Qin:2013mta,Cyrol:2017ewj} and when only the leading Dirac covariant, $\gamma_\mu$, is included, the lack of strength to produce
sufficient DCSB in the quark mass must be compensated by an enhanced gluon-dressing function. This is the premise of the leading {\em rainbow-ladder\/} truncation scheme
which satisfies the axialvector Ward-Green-Takahashi identity and thus correctly describes chiral symmetry and its breaking pattern~\cite{Bender:1996bb}. 

Typically, one neglects three-gluon interactions in this ``Abelian'' truncation and assuming  the corresponding Ward-Green-Takahashi identity holds, i.e. $Z^f_1 =Z^f_2$, one re-expresses 
the kernel of Eq.~\eqref{DSE} as~\cite{Bloch:2002eq,Binosi:2014aea},
\begin{equation}
  Z^f_1 g^2  D_{\mu\nu} (q) \,  \Gamma_\mu^f (k, p) \ =  \  \left ( Z_2^f \right )^2  \mathcal{G} (q^2) \, D_{\mu\nu}^\mathrm{free} (q) \,  \gamma_\mu \ ,
\label{DSEtrunc}  
\end{equation}
where $D_{\mu\nu}^\mathrm{free}  (q) := \big ( g_{\mu\nu} -  q_\mu q_\nu/q^2 \big )/q^2$, $q=k-p$, is the perturbative gluon propagator in Landau gauge. 
An effective model  coupling,  whose momentum dependence  mirrors that of DSE- and lattice-QCD results and yields successful explanations of numerous hadron 
observables,  is given by the sum of two scale-distinct contributions~\cite{Qin:2011xq}:
\begin{equation}
\frac{\mathcal{G} (q^2)}{q^2} =   \frac{8\pi^2}{\omega^4}  D  e^{- q^2/\omega^2 } 
                             +  \frac{8\pi^2 \gamma_m\, \mathcal{F}(q^2) }{\ln \Big [\tau + \big(1 + q^2/\Lambda_\mathrm{QCD}^2 \big )^{\! 2} \Big ] } \, .
\label{qinchang}
\end{equation}
The first term is an infrared-massive and -finite {\em ansatz\/} for the interaction, where  $\gamma_m = 12/(33 - 2N_f )$, $N_f = 4$, $\Lambda_\mathrm{QCD} = 0.234$~GeV; $\tau = e^2 - 1$; 
and $\mathcal{F}(q^2) = [1 - \exp(-q^2/4m^2_t  ) ]/q^2$, $m_t = 0.5$~GeV.  The parameters $\omega$ and $D$ control the width and strength of the interaction, respectively, yet they
should not be thought of as independent~\cite{Binosi:2014aea}.

An even more dramatic simplification is the  CI ansatz in the DSE, namely the following substitution,
\begin{equation}
   g^2 D_{\mu\nu}(q) \, \Gamma^f_\nu(k,p) \quad \Longrightarrow  \quad \Bigg( \frac{4\pi\alpha_{\rm IR}}{m^2_g} \Bigg )_{\! f} \,  \gamma_\mu  \ ,
\end{equation}
where $m_g \simeq 800$~MeV is a gluon mass scale generated dynamically in QCD\footnote{\, See, e.g., discussion in Ref.~\cite{Boucaud:2011ug}.}, and $\alpha_\mathrm{IR}$  is a 
parameter that determines the interaction strength~\cite{Serna:2017nlr}. With this simplification, the solution to the gap equation becomes, 
\begin{equation}
  S^{-1}_f(p)=  i\gamma\cdot p + M_f \ .
  \label{CImass}
\end{equation}
The constituent quark mass in Eq.~\eqref{CImass} is momentum independent, a feature known from nonrelativistic and relativistic constituent quark models~\cite{Ortega:2016pgg,Ortega:2016hde,
deMelo:2014gea,ElBennich:2012ij,daSilva:2012gf,ElBennich:2008xy,ElBennich:2008qa,Yabusaki:2015dca}. As we shall see, in conjunction with the BSE~\eqref{BSE}, this approximation still produces 
a realistic description of static observables and in particular of the mass spectrum of non-exotic ground-state mesons\footnote{\, This is also true for baryons; see Refs.~\cite{Chen:2012qr,Xu:2015kta,Lu:2017cln}.}. 
We stress this is  not only true for heavy quarkonia but also for the Goldstone bosons, since the CI-BSE in rainbow-ladder truncation satisfies the axialvector Ward-Green-Takahashi identity when
 treated in a symmetry-preserving  manner~\cite{Serna:2017nlr,Serna:2016kdb}.  For a given quantum state, the homogeneous BSE is generally given by,
\begin{equation}
  \big [  \Gamma_{\cal M} (k,P)\big ]_{AB}^{fg} =  \int_q^\Lambda \left[K(k,q,P)\right]_{AC,DB}^{fg}  \, \left[S_f(q_+)\, \Gamma_{\cal M}^{fg} (q,P)\, S_g(q_-)\right]_{CD},
\label{BSE}
\end{equation}
 where $\mathcal M$ represents the Dirac spinor structure of the Bethe-Salpeter amplitude $\Gamma_{\cal M} (k;P)$, $K(q,k;P)$ is the fully amputated quark-antiquark scattering kernel, 
 $[A,B,C,D,E,F]$ collectively denote color and spinor indices and $f,g$ are flavor indices. Moreover, $q$ is the relative quark-antiquark momentum and $q_\pm = q  \pm \eta_\pm P$ 
 with the constraint $\eta_+ + \eta_- = 1$. In a Poincar\'e covariant form of this BSE the solutions do not depend on the choice for $\eta_\pm$.
 
 To obtain the spectrum, one introduces a fictitious eigenvalue, $\lambda_n (P^2)$, into the bound-state equation, such that,
\begin{equation}
  \lambda_n (P^2) \, \Gamma_{\cal M}^{fg} (k,P) \ = \int_q^\Lambda  K^{fg}(k,q,P) \, S_f(q_+)\, \Gamma_{\cal M}^{fg} (q,P)\, S_g (q_-) \ ,
\end{equation}
where color and spin indices have been suppressed.  The solution in a particular $J^{PC}$ channel is found for $\lambda_n (M_n^2) - 1=0$ which determines the mass, $M_n$, 
of the {\em n}th bound state with physical solutions for $\lambda_0(M_0^2) = \lambda_1(M_1^2) = \lambda_2(M_2^2)  =  \cdots =  \lambda_i(M_i^2) =1$. Here, the solution for $P^2=-M_0^2$ 
is associated with the hadron's ground-state mass and $M_i$ are the masses of radial excitations ordered as, $M_i < M_{i+1}$,  and $\lambda_0(M_i^2) >  \lambda_1(M_i^2) > \cdots >  
\lambda_i(M_i^2)$~\cite{El-Bennich:2016qmb,El-Bennich:2015kja}.  

\begin{table}[t!]
\def\arraystretch{1.3}

\centering
\begin{tabular}{c|ccccccc}
  \hline  \hline 
  $m_{0^{-(+)}}$  & CI~\cite{Raya:2017ggu} &  CI~\cite{Serna:2017nlr}   &  DSE~\cite{Rojas:2014aka,Mojica:2017tvh} & DSE~\cite{Fischer:2014cfa}  & DSE~\cite{Ding:2015rkn}  
 & DSE~\cite{Hilger:2017jti}  & PDG~\cite{Patrignani:2016xqp} \\ \hline
  $D$  & --- & 1.869  & 2.115  & --- & --- &  1.868 & 1.864  \\
  $D_s$  & --- & 1.977 & 2.130  & --- & --- &  1.872 & 1.968  \\
  $\eta_c(1S)$  &  2.950   &  ---   &  3.065   &  2.925  &  2.980 & 2.672 &   2.983 \\ 
  $\eta_c(2S)$   & --- & ---  &  3.784 & 3.684  & --- & 3.256  & 3.639   \\
  $\eta_b(1S)$  & 9.345 &  --- & ---  & 9.414 & 9.390 &  9.424 & 9.399  \\
  $\eta_b(2S)$   & --- & ---  &  ---  & 9.987  & --- & 9.820  & 9.999   \\
  \hline
  $m_{1^{-(-)}}$  & CI~\cite{Raya:2017ggu} &  CI~\cite{Serna:2017nlr}   &  DSE~\cite{Rojas:2014aka,Mojica:2017tvh} & DSE~\cite{Fischer:2014cfa}  & DSE~\cite{Ding:2015rkn}  
 & DSE~\cite{Hilger:2017jti}  & PDG~\cite{Patrignani:2016xqp} \\ \hline
  $D^*$  & --- & 2.011  & ---  & --- & --- &  --- & 2.010  \\
  $D_s^*$  & --- & 2.098 & ---  & --- & --- &  --- & 2.112  \\
  $J/\psi$   & 3.129  & ---  &  3.114  & 3.113 & 3.070  &  2.840 & 3.097 \\
  $\psi (2S) $   & ---  & ---  &  3.760  & 3.676 & --- & 3.294 & 3.686 \\
  $\Upsilon (1S)$  & 9.460 &  --- & 9.634 & 9.490 & 9.460 & 9.463 & 9.460  \\
  $\Upsilon (2S)$   & --- & ---  &  10.140  & 10.089  & --- & 9.838  & 10.023   \\
  \hline  \hline
\end{tabular}  \par
\caption{Comparative juxtaposition of the charmonium and bottomonium mass spectrum. The $J^{PC}$ meson masses are obtained with different interaction 
models in a rainbow-ladder truncation of Eq.~\eqref{BSE}. Details about the exact form of these interaction ans\"atze are found in the references of each column. 
All masses are in [GeV] and $C$-parity is only a good quantum number for neutral $\bar qq$ eigenstates. Averaged experimental values by the Particle Data 
Group~\cite{Patrignani:2016xqp} are listed  in the last column. } 
\label{table1}
\vspace*{-4mm}
\end{table}

In Table~\ref{table1} we compare a selection of $0^{-+}$ and $1^{--}$ meson masses obtained with the CI and finite-range interaction models in the DSE and BSE kernels of
Eqs.~\eqref{DSE} and \eqref{BSE}. Note that the renormalized running quark mass, $m(\mu) = Z^f_2(\mu,\Lambda)/Z^f_4(\mu,\Lambda) m_f^\mathrm{bm}$ where $Z^f_4$ is 
the renormalization constant in the QCD Lagrangian, can be set to reproduce the experimental $\eta_c$ or $J/\psi$ masses or to minimize the variation of calculated masses 
within a range of experimental values. Therefore, different calculations of, e.g., the $\eta_c(1S)$ lead to variations of the order of 3--5\% except for the value of Ref.~\cite{Hilger:2017jti} 
which is due to their charm current-quark mass, $m_c (19~\mathrm{GeV}) =975$~MeV, using the Alkofer-Watson-Weigel interaction. This value is chosen much smaller, 
$m_c(19~\mathrm{GeV})= 695$~MeV, when the Maris-Tandy model is  employed, yet while the $D$ mass is in very good agreement with experiment, other charmonia masses 
are underestimated. A very similar picture emerges upon examination of the vector-meson states and radial excitations with small variations throughout the models. 
The agreement with data is obvious for the bottomonia with a maximal variation of $\approx 2$--3\%. 

Remarkably, the CI model yields mass values that are consistent with the more sophisticated interaction ans\"atze and in very good agreement with measured values. This
is a feature of the CI --- it is a faithful approximation of the QCD dynamics for static and dynamic observables for $Q^2 \lesssim  M^2_n$~\cite{Bedolla:2016yxq,
Chen:2012qr,Segovia:2013rca}. Notable deviations of predictions of the CI and constituent quark models from measurements can be tested in elastic and transition form 
factors at large $Q^2$~\cite{Cloet:2008re,Chang:2009ae}. Finally, we observe that, so far, none of the BSE truncations provides a solution for the $B$ mesons.

\section{Weak decay constants of ground states and radial excitations}

An  important observable in the context of flavor physics and $CP$ violation is the weak decay constant of mesons.  The axial-vector Ward-Green-Takahashi identity implies 
a defining relation between the kernels in Eqs.~\eqref{DSE} and \eqref{BSE} for pseudoscalar mesons. The identity can be used to prove a  series of Goldberger-Treiman type 
of relations~\cite{Maris:1997hd}, the most important of which relates the two-body problem to the one-body scalar amplitude in the chiral limit ($\mathcal{M} := 0^-$):
\begin{equation}
  f_{0^-}^0 E_{0^-}(k;0)  \ =  \  B^0 (k^2)
\label{GTrelation}
\end{equation}
Here, $E_{0^-}(k;0) $ is the leading covariant of the pseudoscalar meson's Bethe-Salpeter amplitude and the superscript denotes the chiral limit of the quark 
propagator's scalar piece. The weak decay constant, $f_{0^-}$, is given by,
\begin{equation}
  f_{0^-} P_\mu = \langle 0\, |\, \bar q_f \gamma_5 \gamma_\mu q_g \, | \,0^-\rangle  \ = \ \left (Z_2^f Z_2^g \right )^{\frac{1}{2}}  \mathrm{tr}_\mathrm{CD} \!
                                     \int_q^\Lambda \!\! i\,\gamma_5\gamma_\mu \, S_f (q_+)\, \Gamma_{0^-}(q;P)\, S_g (q_-) \ 
\label{fpigen}
\end{equation}
where the trace is over Dirac and color indices.

While the chiral relation \eqref{GTrelation} does not hold anymore for heavy-light mesons, it nevertheless is crucial to preserve the chiral properties of the light quark 
in calculating their masses and weak decay constants. Typically, studies that fail to implement the DCSB properties of light quarks in these systems 
use quark propagators with momentum-independent constituent-quark masses, $M_{u,d} \simeq 250-350$~MeV~\cite{deMelo:2014gea,ElBennich:2012ij,daSilva:2012gf,
ElBennich:2008xy,ElBennich:2008qa,Yabusaki:2015dca}. This leads  to considerable model dependence of heavy-light form factors and couplings~\cite{ElBennich:2009vx}.

An interesting feature is that at first  the weak decay constants of light $0^-$ and $1^-$ mesons increase with the light current-quark mass~\cite{Maris:2005tt}, 
yet tend to level off between the strange- and charm-quark mass. For heavy-heavy and heavy-light mesons the decay constants fall off with increasing meson 
mass as $f_\mathcal{M}  \propto 1/\sqrt{m_\mathcal{M}}$  which is consistent with heavy-quark effective field theory~\cite{Manohar:2000dt}.

\begin{table}[t!]
\def\arraystretch{1.3}
\centering
\begin{tabular}{c|cccccc}
  \hline  \hline 
  $f_{0^{-(+)}}$  & CI~\cite{Raya:2017ggu} &  CI~\cite{Serna:2017nlr}   &  DSE~\cite{Rojas:2014aka,Mojica:2017tvh} &  DSE~\cite{Ding:2015rkn}  
 & DSE~\cite{Hilger:2017jti}  &  Reference  \\ \hline
  $D$  & --- & 0.207  & 0.204 & --- &  0.323 &  0.204  \\
  $D_s$  & --- & 0.240 & 0.249 & --- & 0.269 &  0.258  \\
  $\eta_c(1S)$  & 0.360 &  ---     &  0.389  & 0.371  & 0.322 &  0.361  \\ 
  $\eta_c(2S)$   & --- & ---  &  0.105   & --- & 0.137 &  ---  \\
  $\eta_b(1S)$  & 0.781 &  --- & ---  & 0.768 &  0.378  &  ---  \\
  $\eta_b(2S)$   & --- & ---  &  ---   & --- & 0.263  & ---   \\
  \hline
  $f_{1^{-(-)}}$  & CI~\cite{Raya:2017ggu} &  CI~\cite{Serna:2017nlr}   &  DSE~\cite{Rojas:2014aka,Mojica:2017tvh}  & DSE~\cite{Ding:2015rkn}  
 & DSE~\cite{Hilger:2017jti}  & Reference  \\ \hline
  $D^*$  & --- & 0.281  & ---  &  --- &  --- & 0.278   \\
  $D_s^*$  & --- & 0.276 & ---  &  --- &  --- & 0.322  \\
  $J/\psi$   & 0.291  & ---  &  0.433  & 0.361 & 0.383 &  0.416  \\
  $\psi (2S) $   & ---  & ---  & 0.176 & --- & 0.031 & 0.295 \\
  $\Upsilon (1S)$  & 0.310 &  --- & ---  & 0.666  & 0.393 & 0.715  \\
  $\Upsilon (2S)$   & --- & ---  &  0.564    & --- & 0.283  &  0.497  \\
  \hline  \hline
\end{tabular}  \par
\caption{ Comparison of weak decay constants of $0^{-}$ and $1^{-}$ $D_{(s)}$ mesons and quarkonia in CI and DSE-BSE approaches.  All decay constant values are in [GeV]
and the original values of Refs.~\cite{Ding:2015rkn,Raya:2017ggu} were rescaled by $\surd{2}$. The reference values for quarkonia in the last column are inferred from data~\cite{Patrignani:2016xqp} 
on $^1S_0 \to  \gamma\gamma$  and $^3 S_1 \to e^+ e^-$ decays; for the $D$ and $D_s$ mesons they are extracted from the measurement of $f_{D^+} |V_{cd} |$ and$f_{D^+_s} |V_{cs} |$, 
respectively~\cite{Patrignani:2016xqp};  the $D^*$ and $D_s^*$ decay constants are those of recent lattice-QCD simulations~\cite{Becirevic:2012ti}. }
\label{table2}
\vspace*{-4mm}
\end{table}

This behavior seems to occur as low as $q_f=c$ for $q_f \bar u$ mesons and the inflection point, somewhere above the strange mass, depends on the interaction
ansatz one employs. Since the decay constants depend on the norm of the Bethe-Salpeter amplitudes and are thus proportional to the derivative of the quark propagators, 
it is not surprising that they are more sensitive to the model details than the masses. The transition region between the strange and charm flavors is interesting
in many aspects, as it also describes the mass region where the effects of DCSB and explicit chiral symmetry breaking due to the Higgs mechanism are of the same order. 
A convenient parameter to study the effect of DCSB is the renormalization-point invariant ratio $\zeta := \sigma_f  /M^E_f$  where $\sigma_f $ is defined as a 
constituent-quark $\sigma$ term via the Hellmann-Feynman theorem~\cite{ElBennich:2012tp,Holl:2005st},
\begin{equation}
  \sigma_f :=  m_f(\mu) \frac{\partial M_f^E}{\partial m_f(\mu)} \ ,
\end{equation}
and $M_f^E$ is the definition of the Euclidean constituent-quark mass: $(M^E)^2 := \left \{p^2| p^2 =M^2(p^2) \right \}$. For light quarks the ratio $\zeta$ must vanish as their constituent-quark 
mass owes predominantly to DCSB, whereas for very heavy quarks, $\zeta \to 1$. It turns out that this ratio is $0.5$ for $m_q (19~\mathrm{GeV}) \approx 0.5$~GeV, which is about halfway 
between the strange- and  charm-quark mass at this renormalization scale~\cite{Holl:2005st}. Obviously, where $\zeta \simeq 0.5$ depends on the truncation and interaction model employed 
and should be investigated further. 

The chiral properties of pseudoscalar mesons have consequences for their radially excited states, as the following relation for their weak decay constants holds~\cite{Holl:2004fr},
\begin{equation}
   f_{0^-_n} \, m^2_{0^-_n}  =  2\, m(\mu)\, \rho_{0^-_n} (\mu ) \ ,
 \label{GTrel}  
\end{equation}
with the  pseudoscalar projection of the pseudoscalar meson's Bethe-Salpeter wave function onto the origin in configuration space:
\begin{equation} 
\label{rho}
       \rho_{0^-_n} (\mu )  = -i\, \left (Z_4^f Z_4^g \right )^{\frac{1}{2}} \text{Tr}_\mathrm{CD}\! \int_q^\Lambda\!\!  \gamma_5\,  S_f (q_+)\, \Gamma_{0^-_n}^{fg} (q,P) \, S_g (q_-) \ .
\end{equation}
In QCD, the ultraviolet behaviour of the quark-antiquark scattering kernel and the fact that the Bethe-Salpter wave function $\chi_{0^-_n}(k,P) = S_f (k_+)\, \Gamma_{0^-_n}^{fg}(k,P)\,S_g (k_-)$
is a finite matrix-valued function ensures that  $\rho_{0^-_n} (\mu )$ is finite and does not vanish in the chiral limit~\cite{Maris:1997hd}. Since in this limit the ground-state 
pseudoscalar decay constant is  also characterized by $0 < f_{0^-}^0 < \infty$, it follows that $m_{0^-} = 0$ as expected for a Goldstone boson. Now, we turn our attention to
the radially excited states. By assumption, $m_{0^-_n} > m_{0^-}$ and thus $m_{0^-_n} \neq 0$ for $n>0$ in the chiral limit, and one deduces from  Eq.~\eqref{GTrel} that 
$f_{0^-_n} \equiv 0$, $\forall n>0$. Thus, in the chiral limit the probability for a radially excited state to decay via an electroweak current vanishes. Obviously, the decay constants
of the first radial excitation with heavier quarks are not zero; it was noted that [in GeV]: $f_{\pi(1300)} = -8.3 \times 10^{-4}$; $f_{K (1460)} = -0.017$; $f_{(\bar ss)_1} = -0.0216$; and 
$f_{\eta_c(2S)} = 0.105$. Here again, the weak decay constants are slightly negative near and beyond the chiral limit but reach a turning point in between the $s$- and $c$-quark mass. 

We conclude this section with a selective comparison of weak decay constants for charmed mesons, charmonia and their radially excited states obtained in different approaches
in Table~\ref{table2}. Comparing decay constants of radially excited states with those of their ground states shows that the former remain suppressed even at the charm-mass scale. 
We remark that theoretical values for $f_{\eta_c}$ and $f_{J/\psi}$ vary considerably, up to $\approx30$\% in the case of the $J/\psi$. This reflects again the dependence of the decay 
constants on the details of the interaction. As a generalization one may add that reproducing the mass spectrum of the lower-angular momentum quarkonia  is not difficult --- capturing 
their internal dynamics and electroweak properties is a more challenging task~\cite{Raya:2016yuj}, which is exacerbated in unequal mass systems. This mirrors our experience with light  
hadrons whose mass spectrum is easier to understand than their space- and time-like electromagnetic form factors~\cite{Cloet:2013jya,Eichmann:2016yit,Horn:2016rip,Aznauryan:2012ba,Chen:2017pse}.

\section{Final remarks}

Flavor physics has entered a new era with the plethora of high-precision luminosity measurements at LHC$b$ and provides a fertile playground for {\em indirect\/} tests 
of the Standard Model {\em as well as\/} an opportune field for complementary nonperturbative QCD studies in charm and beauty mesons. Much of our current knowledge about 
the flavor sector and the weak mixing angles is an important guideline to interpret weak interactions of any new possible particle, yet nonperturbative effects are still the major 
source of uncertainty in extracting electroweak phases. Thus, what are seemingly two unrelated fields, flavor and hadron physics, share in common the  difficulties of what is 
frequently called {\em soft physics}. 

Quarkonia, on the other hand, offer additional insights into bound-state dynamics, notably in the domain of two heavy nonrelativistic quarks almost adequately described by a
perturbative interaction. Yet, their rich spectrum defies the simpler picture of quark models with such forms of matter as quark-gluon hybrids, mesonic molecules and tetraquarks. 

In concluding, let us consider the example of a weak $B$-meson decay into two daughter mesons, e.g. $B(p_1) \to \pi(p_2) \pi(q)$, $q=p_2-p_1$. In QCD 
factorization~\cite{Beneke:1999br}, the weak decay amplitude, up to corrections of order $\Lambda_\mathrm{QCD}/m_b$, is expressed as, 
\begin{equation}
  \langle \pi(p_2)\, \pi(q) | Q_i (\mu) | B(p_1) \rangle = F_0^{B\to \pi}\! \int_0^1 \! T_i^I(x)\,  \phi_\pi (x) \,dx  \ + \int_0^1 \! T_i^{II} (\xi,x,y)\, \phi_B (\xi) \phi_\pi (x) \phi_\pi (y) \, d\xi dx dy \ , 
\label{qcdf}
\end{equation}
wherein $Q_i (\mu)$ is a dimension-six four-quark operator of the effective weak Hamiltonian\footnote{\, The corresponding Wilson coefficient, $C_i(\mu)$, in the operator-product 
expansion has been omitted here.} taken at the renormalization scale  $\mu$,  $T_i^{I,II}$ are perturbative hard-scattering kernels, $F_0^{B\to \pi}$ is the scalar heavy-to-light transition form 
factor and $\phi_B$ and $\phi_\pi$ are leading-twist light-front distribution amplitudes. It plainly follows from Eq.~\eqref{qcdf} that the nonperturbative contributions to these amplitudes 
are encoded in form factors and partonic distribution amplitudes. For the latter, it has often been assumed that for any scale $\mu$ the asymptotic limit, $\phi^\mathrm{asy}( x) = 6x(1-x)$, 
is a good approximation. At best, the first few terms of their Gegenbauer expansion, when known, are included. However, as demonstrated in Ref.~\cite{Chang:2013pq}, $\phi^\mathrm{asy}( x)$
is a poor approximation of the complete distribution amplitude even at the scales currently accessible to experiments. It follows that in weak decays of heavy mesons and quarkonia it is advisable
to use the broader $\phi(x)$ amplitudes computed from a large numbers of moments in Refs.~\cite{Ding:2015rkn,Chang:2013pq,Segovia:2013eca,Li:2016mah}. 

This is just one example of how recent developments in continuum QCD approaches within the DSE-BSE framework can be very profitable to precision calculation of weak heavy-meson decays. 
Likewise, the time-like transition form factors, $F^{D(B)\to \pi}(q^2)$, are extracted from complicated hadronic matrix elements for which progress is underway~\cite{Lubicz:2017syv}, 
yet more effort is essential; see the discussion in Ref.~\cite{ElBennich:2009vx}. These examples, amongst many others, and the discussions in the preceding sections serve to highlight
common difficulties and issues faced in hadron and flavor physics. Many theoretical and computational problems of flavored observables will only be overcome with more sophisticated
DSE and BSE kernels. This, on the other hand, provides hadron physicists with challenges and opportunities in the years to come.

\subsection*{Acknowledgements}

This overview is based on the presentation at the ``XLVII International Symposium on Multiparticle Dynamics'' (ISMD), Tlaxcala, Mexico and on related talks at the ECT* workshop 
``The  Charm and Beauty of Strong Interactions" in Trento, Italy, and during the ``Hadrons and Their Properties as a Problem in Strong QCD" workshop at Peking University.
The hospitality of the ISMD organizers and friendly atmosphere in Tlaxcala were strongly appreciated. Work supported by the Brazilian research agencies FAPESP and CNPq
under grant nos. 2016/03154-7 and 458371/2014-9, respectively.

\end{document}